\documentclass[prl,showpacs,preprintnumbers,amsmath,amssymb, superscriptaddress]{revtex4-2}

\usepackage[makeroom]{cancel}

\usepackage{amsmath}    
\usepackage{amssymb}
\usepackage{graphicx}   
\usepackage{verbatim}   
\usepackage{color}      
\usepackage{hyperref}   
\usepackage[normalem]{ulem}
\usepackage{natbib}
\usepackage{enumitem}
\usepackage[displaymath, mathlines,running]{lineno}

\hypersetup{colorlinks,linkcolor=blue,urlcolor=blue,citecolor=blue}

\definecolor{darkgreen}{rgb}{0,0.5,0}
\definecolor{purple}{rgb}{0.35,0,0.35}
\definecolor{orange}{rgb}{1,0.5,0}
\definecolor{darkred}{rgb}{.7,0,0}
\definecolor{darkblue}{rgb}{0,0,.3}
\definecolor{grey}{rgb}{.6,.6,.6}
\definecolor{dimgreen}{rgb}{0.2,0.6,0.1}

\begin{document}

\newcommand{\kv}[0]{\mathbf{k}}
\newcommand{\Rv}[0]{\mathbf{R}}
\newcommand{\Hv}[0]{\mathbf{H}}
\newcommand{\Mv}[0]{\mathbf{M}}
\newcommand{\Vv}[0]{\mathbf{V}}
\newcommand{\Uv}[0]{\mathbf{U}}
\newcommand{\rv}[0]{\mathbf{r}}
\newcommand{\gv}[0]{\mathbf{g}}
\newcommand{\al}[0]{\mathbf{a_{1}}}
\newcommand{\as}[0]{\mathbf{a_{2}}}
\newcommand{\K}[0]{\mathbf{K}}
\newcommand{\Kp}[0]{\mathbf{K'}}
\newcommand{\dkv}[0]{\delta\kv}
\newcommand{\dkx}[0]{\delta k_{x}}
\newcommand{\dky}[0]{\delta k_{y}}
\newcommand{\dk}[0]{\delta k}
\newcommand{\cv}[0]{\mathbf{c}}
\newcommand{\dv}[0]{\mathbf{d}}
\newcommand{\qv}[0]{\mathbf{q}}
\newcommand{\Tr}[0]{\mathrm{Tr}}
\newcommand{\Gv}[0]{\mathbf{G}}
\newcommand{\ev}[0]{\mathbf{e}}
\newcommand{\uu}[1]{\underline{\underline{#1}}}
\newcommand{\ket}[1]{|#1\rangle}
\newcommand{\bra}[1]{\langle#1|}
\newcommand{\average}[1]{\langle #1 \rangle}
\newcommand{\scrap}[1]{{\color{orange}{\sout{#1}}}}
\newcommand{\red}[1]{{\color{red}#1}}
\newcommand{\be}{\begin{equation}}
\newcommand{\ee}{\end{equation}}
\newcommand{\kB}[0]{k_\mathrm{B}}

\newcommand{\jav}[1]{#1}

\title{Kibble-Zurek scaling due to environment temperature quench in the transverse field Ising model}

\author{\'Ad\'am B\'acsi}
\affiliation{MTA-BME Lend\"ulet Topology and Correlation Research Group,
Budapest University of Technology and Economics,  M\H uegyetem rkp. 3., H-1111 Budapest, Hungary}
\affiliation{Department of Mathematics and Computational Sciences, Sz\'echenyi Istv\'an University, 9026 Gy\H or, Hungary}
\affiliation{Jo\v zef Stefan Institute, Jamova 39, Ljubljana SI-1000, Slovenia}
\author{Bal\'azs D\'ora}
\affiliation{MTA-BME Lend\"ulet Topology and Correlation Research Group,
Budapest University of Technology and Economics,  M\H uegyetem rkp. 3., H-1111 Budapest, Hungary}
\affiliation{Department of Theoretical Physics, Institute of Physics, Budapest University of Technology and Economics, M\H uegyetem rkp. 3., H-1111 Budapest, Hungary}
\date{\today}

\begin{abstract}

The Kibble-Zurek mechanism describes defect production due to non-adiabatic passage through a critical point.
Here we study its variant from ramping the environment temperature to a critical point. 
We find that the defect density scales as $\tau^{-d\nu}$ or $\tau^{-d/z}$ for thermal or quantum critical points, respectively,
in terms of the usual critical exponents and $1/\tau$ the speed of the drive.
Both scalings describe reduced defect density compared to conventional Kibble-Zurek mechanism, which stems from the enhanced relaxation due to bath-system interaction.
Ramping to the quantum critical point is investigated by studying the Lindblad equation for the transverse field Ising chain
in the presence of thermalizing bath, with couplings to environment obeying detailed balance, confirming the predicted scaling.
The von-Neumann or the system-bath entanglement entropy follows the same scaling.
Our results are generalized to a large class of dissipative systems with power-law energy dependent bath spectral densities as well.
\end{abstract}

\maketitle

\section{Introduction}

Non-adiabatic dynamics and quantum quenches have been investigated intensively both experimentally and theoretically\cite{polkovnikovrmp,dziarmagareview}.
This allows us to address fundamental questions such as thermalization and equilibration, to introduce non-equilibrium quantum fluctuation
relations\cite{rmptalkner}, to analyze non-linear response. The most archetypical feature  is the Kibble-Zurek mechanism\cite{kibble,zurek,dziarmaga},
which describes universal features of defect production for near adiabatic passages across quantum critical 
points\cite{polkovnikovrmp,dziarmagareview,polkovnikov,keesling,beugnon2017,ko19,PRXQ,damski2008,bialonczyk,qiuscience,cui2020}.
This theory finds application in diverse fields of physics, ranging from quantum and statistical mechanics through cosmology 
and cold atomic systems to condensed matter physics.

The basic idea behind Kibble-Zurek theory is that when a system is driven to\cite{damski2008,bialonczyk} or through\cite{dziarmaga,polkovnikov} the quantum critical point (QCP) by ramping some control parameter,
it undergoes an adiabatic-diabatic transition\cite{campo2011}. In the adiabatic phase, the system has enough time to adjust itself to the new thermodynamic conditions, therefore follows its equilibrium state and the defect production is negligible.
On the other hand, upon entering into the diabatic regime, the relaxation time of the system is longer than the timescale associated to the drive. Therefore, the system cannot adjust itself to new equilibrium conditions and defects are inevitably produced.
The density of defects depends on the rate of change of the control parameter and certain equilibrium critical exponents. 

So far, the Kibble-Zurek mechanism has been exhaustively investigated in closed quantum systems. Recently,
there is a surge of interest towards open quantum systems and non-hermitian
Hamiltonians\cite{ElGanainy2018,rotter,bender98,ashidareview,bergholtz2021,nalbach2015,arceci2018,oshiyama2022}.
These focus on open quantum systems, where dissipation and decoherence through  gain and loss and Lindblad dynamics take place. 
In addition, the Lindblad equation opens the door to study thermalization dynamics by incorporating the principle of detailed 
balance in the couplings to the environment\cite{rajagopal,carmichael,breuer,palmero,reichental}.
Various aspects of the Kibble-Zurek idea has been discussed under 
dissipative conditions\cite{rossini2020,kuo2021,fan2014,Keck,zamora,larson,patane,yin2016,anglin,witkowska,Liu2020,navon,kastner}.

We generalize the Kibble-Zurek scaling for quantum systems containing a QCP, namely the transverse field Ising chain, 
and coupled to a thermalizing  bath within the Lindblad equation. In this case, the relaxation is dominated by the 
system-bath coupling and not by the intrinsic relaxation scale of the QCP. By ramping down the environment temperature to reach the QCP,
we find that the defect density obeys a universal scaling, distinct from the conventional Kibble-Zurek scenario, 
even when the initial temperature is relatively high. This is attributed to the enhanced relaxation due to bath-system interaction.
The thermodynamic entropy of the systems also follows the same scaling.

\section{Results}

\subsection{Kibble-Zurek scaling through driving the environment temperature}

We review first the \jav{conventional} Kibble-Zurek scaling before generalizing it to thermal and quantum phase transition in open quantum systems. 
We study quenching to the critical point, which satisfies the same scaling as ramping through the critical point\cite{damski2008,bialonczyk}.
The reduced temperature is $\tilde T=T-T_c$ with $T_c$ the  critical temperature, and it is driven to the critical point as a function of time\cite{campo2011}. 
\jav{Here, we use the conventional approach of statistical physics that the system exchanges energy with a large heat bath at temperature T but their interaction 
is negligible\cite{clerkrmp}, e.g. the canonical ensemble. As a result, the temperature can appear in the Hamiltonian 
as a parameter through temperature dependent order parameter, external trapping potential etc., and the system is effectively a closed quantum system from the 
dynamics point of view.}
When the critical point is approached, the adiabatic-diabatic transition occurs when the rate, at which we drive the system through $\tilde T(t)$, becomes 
comparable to the inverse of the relaxation time $\tau_\mathrm{rel}$. 
This follows $\tau_\mathrm{rel}\sim {\tilde T}^{-z\nu}$ with
$z$ and $\nu$ the dynamical critical exponent and the exponent associated to the correlation length\cite{sachdev,cardy,continentino}. 
The adiabatic-diabatic transition occurs when these two inverse timescales become comparable
\begin{gather}
\frac{1}{\tilde T} \left|\frac{d\tilde T}{dt}\right|\sim {\tilde T}^{z\nu}.
\label{eq:kzc}
\end{gather}
We consider linear cooling as $\tilde T(t) = \tilde T_0 (1-t/\tau)$ with $\tilde T_0$ the reduced initial temperature $T_0-T_c$ 
and $\tau^{-1}$ the rate of change.
From Eq.  \eqref{eq:kzc},  the adiabatic-diabatic transition temperature is $\tilde T(t_{tr}) \sim \tau^{-1/(1 + z\nu)}$ at the transition time $t_{tr}$. 
After $t_{tr}$, the system leaves the adiabatic time evolution and defect production takes place.
This temperature governs the scaling properties during the diabatic region.
The correlation length scales\cite{cardy} as $\xi\sim 1/(\tilde T(t_{tr}))^\nu$ and in a $d$-dimensional system, 
the density of defects follows
\begin{gather}
n\sim \xi^{-d}\sim \tau^{-\frac{d\nu}{1+z\nu}}\,.
\label{eq:kz}
\end{gather}
\jav{This equation applies for negligible system-heat  bath interaction. Therefore, 
we now discuss the fate of the Kibble-Zurek scaling in the presence of non-negligible system-environment coupling, namely  in
a genuine open quantum system. 
In this case, the relaxation properties of the system are also influenced and even 
dominated} by the interaction with the environment rather than the internal relaxation processes, \jav{namely the coupling to the environment plays a more important 
role than the intrinsic relaxation time of the system.}
Within a Lindblad description\cite{brenes,weiss,carmichael}, the environment is characterized by an effective spectral density $\gamma$, 
which sets the characteristic damping rate, and possesses a given temperature through the temperature dependent environmental occupation numbers.
\jav{In thermal equilibrium, the system itself exchanges energy with the bath and takes its temperature.}

\jav{In the case of driving the environmental temperature}, the adiabatic-diabatic transition is determined by effective spectral density of the environment 
$\gamma$. A more complicated case of energy dependent spectral density is discussed at the end of this section.
Upon changing the environment temperature, the system temperature also changes. 
The rate of change of the system temperature should be compared to $\gamma$ and \emph{not} to the inherent relaxation time of the system, i.e.
\begin{gather}
\frac{1}{\tilde T} \left|\frac{d\tilde T}{dt}\right|\sim \gamma\,.
\label{eqgamma}
\end{gather}
\jav{We note that the r.h.s. of Eq. \eqref{eqgamma} contains in principle also the intrinsic relaxation \emph{rate} of the system, i.e. $\gamma + {\tilde T}^{z\nu}$. However, close to the critical point, the constant environmental coupling $\gamma$ overwhelms
the vanishing intrinsic relaxation rate $T^{z\nu}$ of the system. In other words, the system relaxes through the faster relaxation channel from the environment (if present) rather than the increasingly long intrinsic relaxation time.}
For linear cooling, the adiabatic-diabatic transition happens at time $1/\gamma$ before the critical point is reached. 
The temperature at this time instant is $\tilde T(t_{tr})=\tilde T_0/\gamma\tau$. 
At the scale $\tilde{T}(t_{tr})$, the system crosses over from a mainly adiabatic time evolution, when the density matrix closely follows the equilibrium state, to a diabatic time evolution with significant defect production.
The correlation length scales with this temperature as $\xi\sim 1/(\tilde T(t_{tr}))^{\nu}$ and
the defect density with respect to the thermal expectation value is 
\begin{gather}
n_{thermal}\sim \xi^{-d} \sim \left({\gamma\tau}/{\tilde T_0}\right)^{-d\nu}
\label{eq:kznew}
\end{gather}
for $\gamma\tau\gg 1$.
This applies to thermal phase transitions, driven by $\tilde T$, \jav{in the presence of a finite coupling to environment $\gamma$. In the limit 
of negligible coupling to environment, one has to consider
the intrinsic relaxation time of the system instead, as discussed below Eq. \eqref{eqgamma}, yielding Eq. \eqref{eq:kzc}.} For a quantum phase transition, which occurs at $T_c=0$, however, 
the temperature itself does \emph{not} drive the quantum phase
transition, and the associated thermal correlation length\cite{sachdev,herbutbook} scales as $\xi_T\sim T^{-1/z}$.
Then, 
Eq. \eqref{eq:kznew} is modified for a QCP as
\begin{gather}
n_{qcp}\sim \xi_T^{-d} \sim \left({\gamma\tau}/{T_0}\right)^{-d/z}
\label{eq:kznewq}
\end{gather}
using again the temperature $T(t_{tr})=T_0/\gamma\tau$ at the adiabatic-diabatic transition.
For a given $\tau$, 
the defect density in Eqs. \eqref{eq:kznew} and \eqref{eq:kznewq} is suppressed compared to the conventional Kibble-Zurek case due to the larger exponent. 
The lower defect density is the consequence of the
enhanced relaxation stemming from the bath-system interaction compared to the diverging relaxation time (and vanishing energy scale) for closed quantum systems.
In addition, Eqs. \eqref{eq:kznew} and \eqref{eq:kznewq} predict not only the $\tau$, but also the $T_0$ and $\gamma$ dependence of the
defect density.

We can further generalize these scalings for an environment\cite{weiss,carmichael,breuer} with energy dependent effective spectral density $\gamma(E)\sim |E|^s$
with $s>0$ exponent. The $s=1$ case corresponds to the common Ohmic bath\cite{weiss}.
We find that while Eq. \eqref{eq:kznew} remains unchanged, Eq. \eqref{eq:kznewq} is modified as 
\begin{gather}
n_{qcp}\sim \tau^{-\frac{d}{z(1+s)}}.
\label{kzqnew}
\end{gather}
This follows from realizing that at temperature $T$, the dominant  contribution to damping\cite{brenes}
from environment comes from the $E\sim T$ states, therefore the r.h.s of Eq. \eqref{eqgamma}
becomes $T^s$ through the energy dependent $\gamma$.
Therefore, Eq. \eqref{eq:kznew} remains unchanged since $T_c^s$ is non-singular for any $T_c>0$.
On the other hand, for the quantum case with $T_c=0$, we can realize that  $\frac{1}{\tilde T} \left|\frac{d\tilde T}{dt}\right|\sim T^s$ becomes
 similar to the conventional Kibble-Zurek relation in Eq. \eqref{eq:kzc} with the $z\nu\rightarrow s$ and $\tilde T\rightarrow T $ replacements.
As a result, Eq. \eqref{eq:kznewq} for the
 number of defects after driving the environment temperature to QCP is altered to Eq. \eqref{kzqnew} for a power-law spectral density.
 
We also briefly address the case of non-linear ramps, i.e., when the temperature reaches zero according to $T(t) = T_0(1-t/\tau)^p$. Following the same scaling arguments presented above, the exponent of Eq. \eqref{eq:kz} is modified to $-p\nu d/(1+pz\nu)$ in accordance with Refs. \cite{polkovnikov2008,sen2008}. 
In Eq. \eqref{eq:kznew}, the exponent changes to $-p\nu d$, while in Eq. \eqref{eq:kznewq} and \eqref{kzqnew} the exponents are modified to $-pd/z$ and $-pd/(z(1+ps))$, respectively.
\jav{Further generalizations are also possible for a time dependent coupling, i.e. $\gamma(t)$ as in Ref. \cite{ROSSINI2021}, which is beyond the scope of the present investigation.}


\subsection{Transverse field Ising chain}
The paradigmatic example of a quantum phase transition is represented by the one-dimensional transverse field Ising model\cite{sachdev,dziarmaga,dutta2015,coldea,kinross}  \cite{King2022,Bando2020} .
We demonstrate how the scaling behaviour in Eq. \eqref{eq:kznewq} emerges explicitly in a system whose dynamics is governed by the Lindblad equation.
The model is described by the Hamiltonian Ising coupled spins in a transverse magnetic field as

\begin{gather}
H=-J\sum_{j}\left(g \sigma^x_j + \sigma^z_j \sigma^z_{j+1}\right),
\end{gather}
where $j$ runs over the sites of the one-dimensional chain and $J>0$. The number of sites is $N$ and the length of the chain is $L=Na$ with $a$ the lattice constant. The dimensionless coupling $g>0$ measures the strength of the transverse field.
With a Jordan-Wigner transformation (see Methods), Fourier transformation to momentum space and a Bogoliubov transformation,
the Hamiltonian reduces to
$H = \sum_{k>0,m=\pm}E_k \left(d_{km}^+d_{km}-\frac{1}{2}\right)$,
where $E_k = 2J\sqrt{(g-\cos(ka))^2 + \sin^2(ka)}$
is the energy spectrum of the fermionic excitations and $d_{km}$ are fermionic operators. In the Hamiltonian, the sum runs over the wavenumbers $k=(2n+1)\pi/L$ with an integer $n$.
This quantization corresponds to an antiperiodic boundary condition for the fermionic $c$ operators which is in accordance with periodic boundary condition for the spins\cite{dziarmaga}.

\begin{figure}[h!]
\centering
\includegraphics[width=7cm]{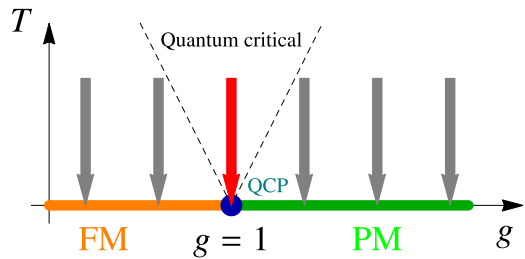}
\caption{Illustration of the phase diagram and the linear cooling protocol, denoted by vertical arrows for the transverse field Ising chain 
at fixed transverse field. The QCP at $g=1$ separates ferro- and paramagnetic phases. }
\label{fig:phdiag}
\end{figure}

The density of states as a function of energy is calculated as
$G(E) = \frac{N}{\pi J}\left(2(g^2 + 1) - (g^2-1)^2\left(\frac{2J}{E}\right)^2  - \left(\frac{E}{2J}\right)^2\right)^{-\frac 12}\,.$
The domain of $G(E)$ is $\Delta<E<2J|g+1|$ with $\Delta = 2J|g-1|$ being the gap.

For $g=1$, the spectrum is gapless and the system realizes a QCP with critical exponents $z=\nu=1$\cite{sachdev} and $d=1$. 
The QCP separates ferromagnetic ($g<1$) and paramagnetic ($g>1$) phases.
In Ref. \cite{dziarmaga}, the quantum quench between the two phases has been studied, i.e., when the system is driven along the $g$ axis through the 
critical point at $g=1$ and zero temperature $T=0$. In the present paper, we consider approaching the QCP from another direction on the phase diagram as illustrated in Fig. \ref{fig:phdiag}.

\subsection{Temperature quench to $T=0$ within the Lindblad equation}
We now couple  the transverse field Ising chain to a thermalizing bath via the Lindblad equation\cite{daley,ashidareview,carmichael,rotter} 
and consider a quench in which the transverse field is kept constant while the \emph{environment} 
temperature is driven linearly from a finite value to zero, see Fig. \ref{fig:phdiag}. 
This yields

\begin{gather}
\partial_t\rho = -i\left[ H,\rho\right] + \sum_{km;\sigma=\uparrow,\downarrow }\gamma_{k,\sigma} \mathcal{D}\left(L_{km,\sigma};\rho; L_{km,\sigma}^+\right),
\label{eq:lind}
\end{gather}
where $\mathcal{D}\left(L;\rho;L^+\right) = L\rho L^+ - \{L^+ L,\rho\}/2$.
In order to thermalize the system, two jump operators are considered for each $k$ and \jav{$m$}, which couple to the eigenstates of the Hamiltonian as $L_{km,\uparrow} = d_{km}^+$
and $L_{km,\downarrow} = d_{km}$, creating and annihilating a fermionic excitation with quantum numbers $m$ and $k$, respectively.
Thermalization is ensured by requiring the couplings to environment to obey detailed balance corresponding to a bath temperature $T$ as 
$\gamma_{k,\downarrow}/\gamma_{k,\uparrow} = e^{\beta E_k}$ with $\beta =1/T$ \cite{breuer}.

Since the goal is to investigate a time-dependent variation of temperature, the temperature dependence of the coupling constants $\gamma_{k,\downarrow}$ and $\gamma_{k,\uparrow}$ is essential. The condition of detailed balance determines their ratio only, while an
 explicit temperature dependence would follow  by performing microscopic derivation of bath correlation functions \cite{breuer,abbruzzo2021}.
For a system of the type in Eq. \eqref{eq:lind}, the temperature dependence can be written as\cite{reichental,abbruzzo2021}

\begin{gather}
\gamma_{k,\downarrow} = \gamma \frac{1}{1 + e^{-\beta E_k}}\qquad \mbox{and} \qquad 
\gamma_{k,\uparrow} = \gamma \frac{1}{1 + e^{\beta E_k}}.
\label{eq:gammas}
\end{gather} 
 The transition rate $\gamma$ is already independent from temperature. 	\jav{We emphasize that the Lindblad master equation makes sense only for small system-bath coupling $\gamma$\cite{breuer,daley,carmichael}.}
We note that the jump operators $L_{km,\sigma}$ are one of many possible choices to describe detailed balance,  
more jump processes between states of different wavenumbers could also be included. 
However, for the sake of simplicity and physical relevance, we focus on the most obvious dissipative processes.

In the followings, we assume that the temperature decreases from an initial temperature $T_0$ to zero as
\begin{gather}
T(t)=T_0\left(1- \frac{t}{\tau}\right), \hspace*{4mm} 0<t<\tau,
\label{eq:lincool}
\end{gather}
and describes linear cooling.
Consequently, through $\beta(t)=1/T(t)$, the coupling constants $\gamma_{k,\uparrow}$ and $\gamma_{k,\downarrow}$ depend on time.
We note that time-dependent coupling as in Eqs. \eqref{eq:gammas} and \eqref{eq:lincool} 
preserve the Markovian approximation leading to the Lindbladian dynamics, as was demonstrated in Refs. \cite{Laine2012,breuertimelocal2019, Donvil2022}: time-local Lindbladians are Markovian as 
long as the coupling constants in Eq. \eqref{eq:gammas} are positive throughout the time evolution, and can be connected to and derived from a suitably  interacting system-environment model.

Assuming thermal equilibrium at $t=0$, the probability that the state $k\jav{m}$ is filled with a fermion is calculated from the Lindblad equation as (see Methods)

\begin{gather}
\langle d_{km}^+d_{km} \rangle \equiv p(E_k,t) = \frac{e^{-\gamma t}}{1 + e^{\beta_0 E_k}} + \gamma\int_0^{t} \frac{e^{-\gamma(t-t')}}{1+e^{\beta(t')E_k}} \mathrm{d}t'
\label{eq:pks}
\end{gather}
which depends on the wavenumber through $E_k$ only.

\subsection{Defect density}
The total density of defects is obtained as

\begin{gather}
n(t)=\frac{1}{N}\sum_{km} p(E_k,t) = \frac{1}{N}\int G(E) p(E,t)\,\mathrm{d}E,
\label{defdens}
\end{gather}
where $G(E)$ is the density of states. The defect density
represents \jav{the number of kinks in the FM state\cite{dziarmaga,sachdev} as the expectation value of $\frac 12\sum_n(1-\sigma^z_n\sigma^z_{n+1})$ or the transverse magnetization in the PM phase, $\sum_n\sigma^x_n$.}
For a perfectly adiabatic quench, the system would reach its ground state and no defects would be present. For finite quench duration, however, a finite number of defects is generated at the end of the quench due to the adiabatic-diabatic transition. 

The final density of defects $n(\tau)$ depends strongly on whether the system is critical or gapped, which influences  the behaviour of the density of states 
at low energies. 
For the critical gapless system ($g=1$), the density of states is constant, $G(E)\sim \mathrm{const.}$ down to $E=0$, while in the gapped phase ($|g-1|\gg 0$), the density of state diverges as $G(E)\sim \sqrt{E/(E-\Delta)}$ at the gap edge, $E\gtrsim\Delta$.

In principle, for the gapped phase, the number of defects after the temperature ramp is expected to be exponentially suppressed on general ground, 
while at or very close to the QCP,
a power-law dependence as in Eq. \eqref{eq:kznewq} is expected.

We start with the behaviour in the gapped phases. The final density of defects is obtained analytically in Methods.
For near-adiabatic quenches in the gapped case ($g$ far from 1) with both $1\ll \beta_0 \Delta$ and $1\ll \gamma\tau$, we find that 

\begin{gather}
n(\tau) =\frac{|1-g|}{\sqrt{2g}}\times\left\{\begin{array}{ll} 
e^{-2\sqrt{\beta_0 \Delta\gamma\tau}},& \beta_0 \Delta\ll \gamma\tau\\
\frac{e^{-\beta_0 \Delta- \gamma\tau}}{\sqrt{\pi \beta_0 \Delta}},& \beta_0 \Delta\gg \gamma\tau
\end{array}\right.
.
\label{eq:gnot1limit}
\end{gather}
In both cases, the number of defects vanishes \jav{faster than power-law} with $\gamma\tau$ for long quenches, in accord with the  gapped 
behaviour of the density of states. \jav{We note that the limit $\gamma\tau\gg 1$ implies large values of $\tau$ but $\gamma$ is kept small to be within the validity of Lindbladian description.}

For the gapless case, $g=1$, the final density of defects follows a power-law dependence as

\begin{gather}
n(\tau) = \frac{\ln 2}{2\pi \beta_0 J} \frac{1 - e^{-\gamma \tau}}{\gamma\tau}\xrightarrow{\gamma\tau\rightarrow\infty}\frac{T_0 \ln 2}{2\pi J } \frac{1}{\gamma\tau}.
\label{eq:g1}
\end{gather}
We note that the same scaling remains valid even if we stop the time evolution at $T_f<T_0$ before reaching $T=0$.
This temperature is reached at $t_f = \tau\left(1-T_f/T_0\right)$ and using the time-dependence of the defect density $n(t)$ from Methods,
we obtain

\begin{gather}
n(t_f) = \frac{\ln 2}{2\pi J\beta_0} 
\left(\frac{T_f}{T_0}   + \frac{1 -e^{-\gamma \tau \left(1-T_f/T_0\right)}}{\gamma\tau}\right).
\end{gather}
The first term describes the defect density in thermal state at $T_f$, while the 
second term comes from the surplus defect density which scales again as $\tau^{-1}$ for long quenches.

\begin{figure}[h]
\centering
\hspace{-0.4cm}\includegraphics[width=9cm]{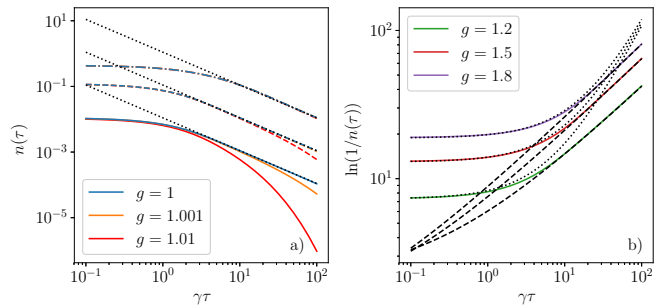}
\caption{Numerical results for the density of defects, $n(\tau)$, at the end of the quench for various values of $g$. 
a) Cooling to or very close to the QCP. The solid, dashed and dash-dotted line corresponds to the initial temperature of $T_0/J=0.1$, 1 and 10, respectively. 
For each initial temperature, the black dotted line shows the $\tau^{-1}$ behavior of Eq. \eqref{eq:g1}. 
b) Far from the QCP, the numerical results (solid lines) agree with Eq. \eqref{eq:gnot1limit} (dashed and dotted lines) in limiting cases, the initial temperature is $T_0/J= 0.1$. For $g<1$, similar $\tau$-dependences are found.}
\label{fig:Nsfinal}
\end{figure}

We have also studied the full lattice version of the model by performing the energy  and the temporal integrals numerically in Eqs. \eqref{eq:pks} and \eqref{defdens}. 
Eqs. \eqref{eq:gnot1limit}-\eqref{eq:g1} agree nicely with  the  numerically exact results in Fig. \ref{fig:Nsfinal} including the $\gamma\tau$ and $T_0$ dependences.
Eq. \eqref{eq:g1} indeed shows the expected power-law decay in the adiabatic limit as Eq. \eqref{eq:kznewq} with $z =d = 1$.
 The exponent of the decay differs from the conventional Kibble-Zurek exponent from Eq. \eqref{eq:kz}, which would predict $n(\tau)\sim \tau^{-1/2}$ 
 for the transverse field Ising
model \cite{dziarmaga}. The difference is 
explained by the bath-system interaction preventing the system from "critical slowing down" and enhancing relaxation throughout the diabatic region. 
We have also checked numerically (see Methods) that our scaling from Eq. \eqref{kzqnew} remains valid in the presence of an Ohmic bath, when the coupling to environment becomes energy dependent as $\gamma\sim |E|^s$ with $s=1$.
In this case, the modified scaling reads as $\tau^{-1/2}$, which is perfectly captured by our exact numerics in Methods.

\subsection{Entropy}
While closed quantum systems are typically described by a wavefunction, open quantum systems possess a density matrix. This allows us to calculate the thermodynamic entropy after the temperature ramp, which also quantifies the entanglement between the system of interest and its environment.
The entropy change represents a useful measure of the adiabaticity of the quench. For the transverse field Ising model, after a perfectly 
adiabatic quench, the system is expected to reach the pure ground state with vanishing entropy.
For a quench of finite duration, however, some entropy is unavoidably generated. The irreversible 
entropy production from Kibble-Zurek theory was touched upon in Ref. \cite{deffner17}.
Based on the occupation probabilities $p_{ks}(t)$ in Eq. \eqref{eq:pks}, the entropy is

\begin{gather}
S(t) = -\sum_{km}\left(p(E_k,t) \ln p(E_k,t) + (1-p(E_k,t))\ln(1-p(E_k,t))\right).
\label{eq:entropy0}
\end{gather}
The final entropy, $S(\tau)$, depends on the final occupation probabilities $p(E_k,\tau)$. 

\begin{figure}[h]
\centering
\hspace{-0.3cm}\includegraphics[width=9cm]{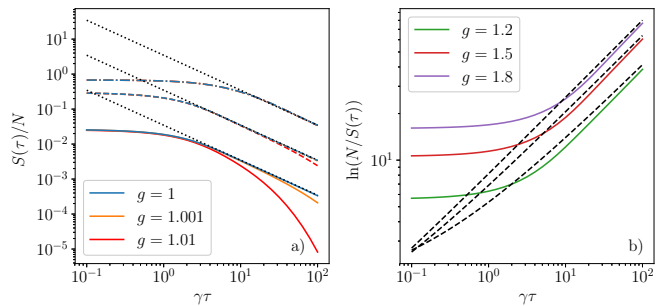}
\caption{Numerical results for the entropy, $S(\tau)$, at the end of the quench for various values of $g$. a) 
Cooling to or very close to the QCP. The solid, dashed and dash-dotted line corresponds to the initial temperature of $T_0/J =0.1$, 1 and 10, respectively. 
For each initial temperature, the black dotted line shows the $\tau^{-1}$ behavior of Eq. \eqref{eq:finent}. b) Numerical results far from the QCP with the 
initial temperature $T_0/J =0.1$, the black dashed lines denote $2Nn(\tau)$.}
\label{fig:finent}
\end{figure}

In the gapped phase with $1\ll\beta_0\Delta \ll\gamma\tau$, the final entropy is obtained 
as $S(\tau) \approx  2 N n(\tau)$,
displayed in Fig. \ref{fig:finent}.
Similarly to the defect density, the most interesting case involves quench to the gapless, critical system with $g=1$.
For long quenches, 
we get

\begin{gather}
S(\tau) \approx \frac{N T_0 \ln 2}{2 J  \gamma \tau}\,.
\label{eq:finent}
\end{gather}
The final entropy is plotted in Fig. \ref{fig:finent} and is in good agreement with the numerical result for long quenches. For quenches terminating 
away from the QCP, the entropy gets exponentially suppressed in $\tau$, similarly to the defect density.
We expect that the residual entropy should scale as $S(T=T_0/\gamma\tau)$ in general, where $S(T)$ is the equilibrium thermal entropy of the system. 
The temperature $T_0/\gamma\tau$ gets imprinted into the dynamics of the system at the  adiabatic-diabatic transition.
For the transverse field Ising chain, the thermal entropy scales\cite{liang} as $S(T)\sim T$, which explains the scaling in Eq. \eqref{eq:finent}.

\section{Discussion}
We have studied the effect of cooling the environment temperature to a thermal or quantum critical point, and its influence on Kibble-Zurek scaling. We find that the diverging relaxation time, associated to critical points, gets replaced by the inverse coupling to environment, which results in a suppressed, but universal scaling of the defect density.
By investigating the dissipative version of the transverse field Ising chain, we verify this prediction and also find that by ramping down the temperature to a gapped ground state, the defect density follows an exponential scaling with the ramp time. The system-bath entanglement entropy follows the same universal 
scaling and should be accessible experimentally,
together with the defect density\cite{keesling}.

For a power-law energy dependent bath spectral function with exponent $s$, the obtained Kibble-Zurek scaling in Eq. \eqref{kzqnew} 
is identical to the conventional one in Eq. \eqref{eq:kz} when $z\nu=1/s$.
Typically these numbers are of order one, therefore one can easily get identical scaling for a temperature quench to the conventional Kibble-Zurek scenario.
In particular, as we demonstrated, an Ohmic bath with $s=1$ for the transverse field Ising chain produces the conventional behaviour for the defect density 
with exponent 1/2.
Experimentally, our results can be tested similarly to  Ref. \cite{PRXQ}.

The universal scaling of the defect density in terms of the quench duration for ramping to a quantum critical point applies 
to a large variety of open quantum systems, ranging from energy independent through  subohmic and ohmic to super ohmic  bath spectral densities.
Not only are these results relevant in highlighting universal features during near-adiabatic cooling processes  but can also be beneficial for
quantum thermodynamics\cite{Alicki2018} for efficient heat pumps or quantum refrigerators\cite{gluza}.
Moreover, 
 understanding defect production through temperature variations close to quantum
critical points promises to be important for smart design of adiabatic quantum computation protocols\cite{nielsen} in open quantum systems.

\begin{acknowledgments}
We thank B. Gul\'acsi for useful discussions.
This research is supported by the National Research, Development and
Innovation Office - NKFIH  within the Quantum Technology National Excellence
Program (Project No.~2017-1.2.1-NKP-2017-00001), K134437,  K142179,  by the BME-Nanotechnology
FIKP grant (BME FIKP-NAT), and by a grant of the Ministry of Research, Innovation and
 Digitization, CNCS/CCCDI-UEFISCDI, under projects number PN-III-P4-ID-PCE-2020-0277.
 \jav{\'A. B. acknowledges the support of the Slovenian Research Agency (ARRS) under J1-3008.}

\end{acknowledgments}




\bibliographystyle{apsrev}
\bibliography{lindblad,wboson1}

\begin{thebibliography}{10}
\expandafter\ifx\csname bibnamefont\endcsname\relax
  \def\bibnamefont#1{#1}\fi
\expandafter\ifx\csname bibfnamefont\endcsname\relax
  \def\bibfnamefont#1{#1}\fi
\expandafter\ifx\csname url\endcsname\relax
  \def\url#1{\texttt{#1}}\fi
\expandafter\ifx\csname urlprefix\endcsname\relax\def\urlprefix{URL }\fi
\providecommand{\bibinfo}[2]{#2}
\providecommand{\eprint}[2][]{\url{#2}}

\bibitem{polkovnikovrmp}
\bibinfo{author}{\bibfnamefont{A.}~\bibnamefont{Polkovnikov}},
  \bibinfo{author}{\bibfnamefont{K.}~\bibnamefont{Sengupta}},
  \bibinfo{author}{\bibfnamefont{A.}~\bibnamefont{Silva}}, \bibnamefont{and}
  \bibinfo{author}{\bibfnamefont{M.}~\bibnamefont{Vengalattore}},
  \emph{\bibinfo{title}{\textit{Colloquium} : Nonequilibrium dynamics of closed
  interacting quantum systems}}, \bibinfo{journal}{Rev. Mod. Phys.}
  \textbf{\bibinfo{volume}{83}}, \bibinfo{pages}{863} (\bibinfo{year}{2011}).

\bibitem{dziarmagareview}
\bibinfo{author}{\bibfnamefont{J.}~\bibnamefont{Dziarmaga}},
  \emph{\bibinfo{title}{Dynamics of a quantum phase transition and relaxation
  to a steady state}}, \bibinfo{journal}{Adv. Phys.}
  \textbf{\bibinfo{volume}{59}}, \bibinfo{pages}{1063} (\bibinfo{year}{2010}).

\bibitem{rmptalkner}
\bibinfo{author}{\bibfnamefont{M.}~\bibnamefont{Campisi}},
  \bibinfo{author}{\bibfnamefont{P.}~\bibnamefont{H\"anggi}}, \bibnamefont{and}
  \bibinfo{author}{\bibfnamefont{P.}~\bibnamefont{Talkner}},
  \emph{\bibinfo{title}{\textit{Colloquium} : Quantum fluctuation relations:
  Foundations and applications}}, \bibinfo{journal}{Rev. Mod. Phys.}
  \textbf{\bibinfo{volume}{83}}, \bibinfo{pages}{771} (\bibinfo{year}{2011}).

\bibitem{kibble}
\bibinfo{author}{\bibfnamefont{T.~W.~B.} \bibnamefont{Kibble}},
  \emph{\bibinfo{title}{Topology of cosmic domains and strings}},
  \bibinfo{journal}{J. Phys. A} \textbf{\bibinfo{volume}{9}},
  \bibinfo{pages}{1387} (\bibinfo{year}{1976}).

\bibitem{zurek}
\bibinfo{author}{\bibfnamefont{W.~H.} \bibnamefont{Zurek}},
  \emph{\bibinfo{title}{Cosmological experiments in superfluid helium?}},
  \bibinfo{journal}{Nature} \textbf{\bibinfo{volume}{317}},
  \bibinfo{pages}{505} (\bibinfo{year}{1985}).

\bibitem{dziarmaga}
\bibinfo{author}{\bibfnamefont{J.}~\bibnamefont{Dziarmaga}},
  \emph{\bibinfo{title}{Dynamics of a quantum phase transition: Exact solution
  of the quantum ising model}}, \bibinfo{journal}{Phys. Rev. Lett.}
  \textbf{\bibinfo{volume}{95}}, \bibinfo{pages}{245701}
  (\bibinfo{year}{2005}).

\bibitem{polkovnikov}
\bibinfo{author}{\bibfnamefont{A.}~\bibnamefont{Polkovnikov}},
  \emph{\bibinfo{title}{Universal adiabatic dynamics in the vicinity of a
  quantum critical point}}, \bibinfo{journal}{Phys. Rev. B}
  \textbf{\bibinfo{volume}{72}}, \bibinfo{pages}{161201}
  (\bibinfo{year}{2005}).

\bibitem{keesling}
\bibinfo{author}{\bibfnamefont{A.}~\bibnamefont{Keesling}},
  \bibinfo{author}{\bibfnamefont{A.}~\bibnamefont{Omran}},
  \bibinfo{author}{\bibfnamefont{H.}~\bibnamefont{Levine}},
  \bibinfo{author}{\bibfnamefont{H.}~\bibnamefont{Bernien}},
  \bibinfo{author}{\bibfnamefont{H.}~\bibnamefont{Pichler}},
  \bibinfo{author}{\bibfnamefont{S.}~\bibnamefont{Choi}},
  \bibinfo{author}{\bibfnamefont{R.}~\bibnamefont{Samajdar}},
  \bibinfo{author}{\bibfnamefont{S.}~\bibnamefont{Schwartz}},
  \bibinfo{author}{\bibfnamefont{P.}~\bibnamefont{Silvi}},
  \bibinfo{author}{\bibfnamefont{S.}~\bibnamefont{Sachdev}},
  \bibinfo{author}{\bibfnamefont{P.}~\bibnamefont{Zoller}},
  \bibinfo{author}{\bibfnamefont{M.}~\bibnamefont{Endres}}, \emph{et~al.},
  \emph{\bibinfo{title}{Quantum kibble-zurek mechanism and critical dynamics on
  a programmable rydberg simulator}}, \bibinfo{journal}{Nature}
  \textbf{\bibinfo{volume}{568}}, \bibinfo{pages}{207} (\bibinfo{year}{2019}).

\bibitem{beugnon2017}
\bibinfo{author}{\bibfnamefont{J.}~\bibnamefont{Beugnon}} \bibnamefont{and}
  \bibinfo{author}{\bibfnamefont{N.}~\bibnamefont{Navon}},
  \emph{\bibinfo{title}{Exploring the kibble{\textendash}zurek mechanism with
  homogeneous bose gases}}, \bibinfo{journal}{Journal of Physics B: Atomic,
  Molecular and Optical Physics}
  \textbf{\bibinfo{volume}{50}}(\bibinfo{number}{2}), \bibinfo{pages}{022002}
  (\bibinfo{year}{2017}).

\bibitem{ko19}
\bibinfo{author}{\bibfnamefont{B.}~\bibnamefont{Ko}},
  \bibinfo{author}{\bibfnamefont{J.~W.} \bibnamefont{Park}}, \bibnamefont{and}
  \bibinfo{author}{\bibfnamefont{Y.}~\bibnamefont{Shin}},
  \emph{\bibinfo{title}{Kibble-zurek universality in a strongly interacting
  fermi superfluid}}, \bibinfo{journal}{Nat. Phys.}
  \textbf{\bibinfo{volume}{15}}, \bibinfo{pages}{1227} (\bibinfo{year}{2019}).

\bibitem{PRXQ}
\bibinfo{author}{\bibfnamefont{L.}~\bibnamefont{Xiao}},
  \bibinfo{author}{\bibfnamefont{D.}~\bibnamefont{Qu}},
  \bibinfo{author}{\bibfnamefont{K.}~\bibnamefont{Wang}},
  \bibinfo{author}{\bibfnamefont{H.-W.} \bibnamefont{Li}},
  \bibinfo{author}{\bibfnamefont{J.-Y.} \bibnamefont{Dai}},
  \bibinfo{author}{\bibfnamefont{B.}~\bibnamefont{D\'ora}},
  \bibinfo{author}{\bibfnamefont{M.}~\bibnamefont{Heyl}},
  \bibinfo{author}{\bibfnamefont{R.}~\bibnamefont{Moessner}},
  \bibinfo{author}{\bibfnamefont{W.}~\bibnamefont{Yi}}, \bibnamefont{and}
  \bibinfo{author}{\bibfnamefont{P.}~\bibnamefont{Xue}},
  \emph{\bibinfo{title}{Non-hermitian kibble-zurek mechanism with tunable
  complexity in single-photon interferometry}}, \bibinfo{journal}{PRX Quantum}
  \textbf{\bibinfo{volume}{2}}, \bibinfo{pages}{020313} (\bibinfo{year}{2021}).

\bibitem{damski2008}
\bibinfo{author}{\bibfnamefont{B.}~\bibnamefont{Damski}} \bibnamefont{and}
  \bibinfo{author}{\bibfnamefont{W.~H.} \bibnamefont{Zurek}},
  \emph{\bibinfo{title}{How to fix a broken symmetry: quantum dynamics of
  symmetry restoration in a ferromagnetic bose{\textendash}einstein
  condensate}}, \bibinfo{journal}{New Journal of Physics}
  \textbf{\bibinfo{volume}{10}}(\bibinfo{number}{4}), \bibinfo{pages}{045023}
  (\bibinfo{year}{2008}).

\bibitem{bialonczyk}
\bibinfo{author}{\bibfnamefont{M.}~\bibnamefont{Bia{\l}o{\'{n}}czyk}}
  \bibnamefont{and} \bibinfo{author}{\bibfnamefont{B.}~\bibnamefont{Damski}},
  \emph{\bibinfo{title}{One-half of the kibble{\textendash}zurek quench
  followed by free evolution}}, \bibinfo{journal}{Journal of Statistical
  Mechanics: Theory and Experiment}
  \textbf{\bibinfo{volume}{2018}}(\bibinfo{number}{7}), \bibinfo{pages}{073105}
  (\bibinfo{year}{2018}).

\bibitem{qiuscience}
\bibinfo{author}{\bibfnamefont{L.-Y.} \bibnamefont{Qiu}},
  \bibinfo{author}{\bibfnamefont{H.-Y.} \bibnamefont{Liang}},
  \bibinfo{author}{\bibfnamefont{Y.-B.} \bibnamefont{Yang}},
  \bibinfo{author}{\bibfnamefont{H.-X.} \bibnamefont{Yang}},
  \bibinfo{author}{\bibfnamefont{T.}~\bibnamefont{Tian}},
  \bibinfo{author}{\bibfnamefont{Y.}~\bibnamefont{Xu}}, \bibnamefont{and}
  \bibinfo{author}{\bibfnamefont{L.-M.} \bibnamefont{Duan}},
  \emph{\bibinfo{title}{Observation of generalized kibble-zurek mechanism
  across a first-order quantum phase transition in a spinor condensate}},
  \bibinfo{journal}{Science Advances}
  \textbf{\bibinfo{volume}{6}}(\bibinfo{number}{21}), \bibinfo{pages}{eaba7292}
  (\bibinfo{year}{2020}).

\bibitem{cui2020}
\bibinfo{author}{\bibfnamefont{J.}~\bibnamefont{Cui}},
  \bibinfo{author}{\bibfnamefont{F.~J.} \bibnamefont{{G\'omez-Ruiz}}},
  \bibinfo{author}{\bibfnamefont{Y.-F.} \bibnamefont{Huang}},
  \bibinfo{author}{\bibfnamefont{C.-F.} \bibnamefont{Li}},
  \bibinfo{author}{\bibfnamefont{G.-C.} \bibnamefont{Guo}}, \bibnamefont{and}
  \bibinfo{author}{\bibfnamefont{A.}~\bibnamefont{{del Campo}}},
  \emph{\bibinfo{title}{Experimentally testing quantum critical dynamics beyond
  the kibble–zurek mechanism}}, \bibinfo{journal}{Commun. Phys.}
  \textbf{\bibinfo{volume}{3}}, \bibinfo{pages}{44} (\bibinfo{year}{2020}).

\bibitem{campo2011}
\bibinfo{author}{\bibfnamefont{A.}~\bibnamefont{del Campo}},
  \bibinfo{author}{\bibfnamefont{A.}~\bibnamefont{Retzker}}, \bibnamefont{and}
  \bibinfo{author}{\bibfnamefont{M.~B.} \bibnamefont{Plenio}},
  \emph{\bibinfo{title}{The inhomogeneous kibble{\textendash}zurek mechanism:
  vortex nucleation during bose{\textendash}einstein condensation}},
  \bibinfo{journal}{New Journal of Physics}
  \textbf{\bibinfo{volume}{13}}(\bibinfo{number}{8}), \bibinfo{pages}{083022}
  (\bibinfo{year}{2011}).

\bibitem{ElGanainy2018}
\bibinfo{author}{\bibfnamefont{R.}~\bibnamefont{El-Ganainy}},
  \bibinfo{author}{\bibfnamefont{K.~G.} \bibnamefont{Makris}},
  \bibinfo{author}{\bibfnamefont{M.}~\bibnamefont{Khajavikhan}},
  \bibinfo{author}{\bibfnamefont{Z.~H.} \bibnamefont{Musslimani}},
  \bibinfo{author}{\bibfnamefont{S.}~\bibnamefont{Rotter}}, \bibnamefont{and}
  \bibinfo{author}{\bibfnamefont{D.~N.} \bibnamefont{Christodoulides}},
  \emph{\bibinfo{title}{Non-hermitian physics and pt symmetry}},
  \bibinfo{journal}{Nat. Phys.}
  \textbf{\bibinfo{volume}{14}}(\bibinfo{number}{1}), \bibinfo{pages}{11}
  (\bibinfo{year}{2018}).

\bibitem{rotter}
\bibinfo{author}{\bibfnamefont{I.}~\bibnamefont{Rotter}} \bibnamefont{and}
  \bibinfo{author}{\bibfnamefont{J.~P.} \bibnamefont{Bird}},
  \emph{\bibinfo{title}{A review of progress in the physics of open quantum
  systems: theory and experiment}}, \bibinfo{journal}{Rep. Prog. Phys.}
  \textbf{\bibinfo{volume}{78}}, \bibinfo{pages}{114001}
  (\bibinfo{year}{2015}).

\bibitem{bender98}
\bibinfo{author}{\bibfnamefont{C.~M.} \bibnamefont{Bender}} \bibnamefont{and}
  \bibinfo{author}{\bibfnamefont{S.}~\bibnamefont{Boettcher}},
  \emph{\bibinfo{title}{Real spectra in non-hermitian hamiltonians having
  $\mathcal{PT}$ symmetry}}, \bibinfo{journal}{Phys. Rev. Lett.}
  \textbf{\bibinfo{volume}{80}}, \bibinfo{pages}{5243} (\bibinfo{year}{1998}).

\bibitem{ashidareview}
\bibinfo{author}{\bibfnamefont{Y.}~\bibnamefont{Ashida}},
  \bibinfo{author}{\bibfnamefont{Z.}~\bibnamefont{Gong}}, \bibnamefont{and}
  \bibinfo{author}{\bibfnamefont{M.}~\bibnamefont{Ueda}},
  \emph{\bibinfo{title}{Non-hermitian physics}}, \bibinfo{journal}{Advances in
  Physics} \textbf{\bibinfo{volume}{69}}, \bibinfo{pages}{3}
  (\bibinfo{year}{2020}).

\bibitem{bergholtz2021}
\bibinfo{author}{\bibfnamefont{E.~J.} \bibnamefont{Bergholtz}},
  \bibinfo{author}{\bibfnamefont{J.~C.} \bibnamefont{Budich}},
  \bibnamefont{and} \bibinfo{author}{\bibfnamefont{F.~K.} \bibnamefont{Kunst}},
  \emph{\bibinfo{title}{Exceptional topology of non-hermitian systems}},
  \bibinfo{journal}{Rev. Mod. Phys.} \textbf{\bibinfo{volume}{93}},
  \bibinfo{pages}{015005} (\bibinfo{year}{2021}).

\bibitem{nalbach2015}
\bibinfo{author}{\bibfnamefont{P.}~\bibnamefont{Nalbach}},
  \bibinfo{author}{\bibfnamefont{S.}~\bibnamefont{Vishveshwara}},
  \bibnamefont{and} \bibinfo{author}{\bibfnamefont{A.~A.} \bibnamefont{Clerk}},
  \emph{\bibinfo{title}{Quantum kibble-zurek physics in the presence of
  spatially correlated dissipation}}, \bibinfo{journal}{Phys. Rev. B}
  \textbf{\bibinfo{volume}{92}}, \bibinfo{pages}{014306}
  (\bibinfo{year}{2015}).

\bibitem{arceci2018}
\bibinfo{author}{\bibfnamefont{L.}~\bibnamefont{Arceci}},
  \bibinfo{author}{\bibfnamefont{S.}~\bibnamefont{Barbarino}},
  \bibinfo{author}{\bibfnamefont{D.}~\bibnamefont{Rossini}}, \bibnamefont{and}
  \bibinfo{author}{\bibfnamefont{G.~E.} \bibnamefont{Santoro}},
  \emph{\bibinfo{title}{Optimal working point in dissipative quantum
  annealing}}, \bibinfo{journal}{Phys. Rev. B} \textbf{\bibinfo{volume}{98}},
  \bibinfo{pages}{064307} (\bibinfo{year}{2018}).

\bibitem{oshiyama2022}
\bibinfo{author}{\bibfnamefont{H.}~\bibnamefont{Oshiyama}},
  \bibinfo{author}{\bibfnamefont{S.}~\bibnamefont{Suzuki}}, \bibnamefont{and}
  \bibinfo{author}{\bibfnamefont{N.}~\bibnamefont{Shibata}},
  \emph{\bibinfo{title}{Classical simulation and theory of quantum annealing in
  a thermal environment}}, \bibinfo{journal}{Phys. Rev. Lett.}
  \textbf{\bibinfo{volume}{128}}, \bibinfo{pages}{170502}
  (\bibinfo{year}{2022}).

\bibitem{rajagopal}
\bibinfo{author}{\bibfnamefont{A.}~\bibnamefont{Rajagopal}},
  \emph{\bibinfo{title}{The principle of detailed balance and the lindblad
  dissipative quantum dynamics}}, \bibinfo{journal}{Physics Letters A}
  \textbf{\bibinfo{volume}{246}}(\bibinfo{number}{3}), \bibinfo{pages}{237}
  (\bibinfo{year}{1998}).

\bibitem{carmichael}
\bibinfo{author}{\bibfnamefont{H.}~\bibnamefont{Carmichael}},
  \emph{\bibinfo{title}{An Open Systems Approach to Quantum Optics}}
  (\bibinfo{publisher}{Springer-Verlag}, \bibinfo{address}{Berlin},
  \bibinfo{year}{1993}).

\bibitem{breuer}
\bibinfo{author}{\bibfnamefont{H.}~\bibnamefont{Breuer}} \bibnamefont{and}
  \bibinfo{author}{\bibfnamefont{F.}~\bibnamefont{Petruccione}},
  \emph{\bibinfo{title}{The Theory of Open Quantum Systems}}
  (\bibinfo{publisher}{Oxford University Press}, \bibinfo{year}{2002}).

\bibitem{palmero}
\bibinfo{author}{\bibfnamefont{M.}~\bibnamefont{Palmero}},
  \bibinfo{author}{\bibfnamefont{X.}~\bibnamefont{Xu}},
  \bibinfo{author}{\bibfnamefont{C.}~\bibnamefont{Guo}}, \bibnamefont{and}
  \bibinfo{author}{\bibfnamefont{D.}~\bibnamefont{Poletti}},
  \emph{\bibinfo{title}{Thermalization with detailed-balanced two-site lindblad
  dissipators}}, \bibinfo{journal}{Phys. Rev. E}
  \textbf{\bibinfo{volume}{100}}, \bibinfo{pages}{022111}
  (\bibinfo{year}{2019}).

\bibitem{reichental}
\bibinfo{author}{\bibfnamefont{I.}~\bibnamefont{Reichental}},
  \bibinfo{author}{\bibfnamefont{A.}~\bibnamefont{Klempner}},
  \bibinfo{author}{\bibfnamefont{Y.}~\bibnamefont{Kafri}}, \bibnamefont{and}
  \bibinfo{author}{\bibfnamefont{D.}~\bibnamefont{Podolsky}},
  \emph{\bibinfo{title}{Thermalization in open quantum systems}},
  \bibinfo{journal}{Phys. Rev. B} \textbf{\bibinfo{volume}{97}},
  \bibinfo{pages}{134301} (\bibinfo{year}{2018}).

\bibitem{rossini2020}
\bibinfo{author}{\bibfnamefont{D.}~\bibnamefont{Rossini}} \bibnamefont{and}
  \bibinfo{author}{\bibfnamefont{E.}~\bibnamefont{Vicari}},
  \emph{\bibinfo{title}{Dynamic kibble-zurek scaling framework for open
  dissipative many-body systems crossing quantum transitions}},
  \bibinfo{journal}{Phys. Rev. Research} \textbf{\bibinfo{volume}{2}},
  \bibinfo{pages}{023211} (\bibinfo{year}{2020}).

\bibitem{kuo2021}
\bibinfo{author}{\bibfnamefont{W.-T.} \bibnamefont{Kuo}},
  \bibinfo{author}{\bibfnamefont{D.}~\bibnamefont{Arovas}},
  \bibinfo{author}{\bibfnamefont{S.}~\bibnamefont{Vishveshwara}}, ,
  \bibnamefont{and} \bibinfo{author}{\bibfnamefont{Y.-Z.} \bibnamefont{You}},
  \emph{\bibinfo{title}{{Decoherent quench dynamics across quantum phase
  transitions}}}, \bibinfo{journal}{SciPost Phys.}
  \textbf{\bibinfo{volume}{11}}, \bibinfo{pages}{84} (\bibinfo{year}{2021}).

\bibitem{fan2014}
\bibinfo{author}{\bibfnamefont{S.}~\bibnamefont{Yin}},
  \bibinfo{author}{\bibfnamefont{P.}~\bibnamefont{Mai}}, \bibnamefont{and}
  \bibinfo{author}{\bibfnamefont{F.}~\bibnamefont{Zhong}},
  \emph{\bibinfo{title}{Nonequilibrium quantum criticality in open systems: The
  dissipation rate as an additional indispensable scaling variable}},
  \bibinfo{journal}{Phys. Rev. B} \textbf{\bibinfo{volume}{89}},
  \bibinfo{pages}{094108} (\bibinfo{year}{2014}).

\bibitem{Keck}
\bibinfo{author}{\bibfnamefont{M.}~\bibnamefont{Keck}},
  \bibinfo{author}{\bibfnamefont{S.}~\bibnamefont{Montangero}},
  \bibinfo{author}{\bibfnamefont{G.~E.} \bibnamefont{Santoro}},
  \bibinfo{author}{\bibfnamefont{R.}~\bibnamefont{Fazio}}, \bibnamefont{and}
  \bibinfo{author}{\bibfnamefont{D.}~\bibnamefont{Rossini}},
  \emph{\bibinfo{title}{Dissipation in adiabatic quantum computers: lessons
  from an exactly solvable model}}, \bibinfo{journal}{New Journal of Physics}
  \textbf{\bibinfo{volume}{19}}(\bibinfo{number}{11}), \bibinfo{pages}{113029}
  (\bibinfo{year}{2017}).

\bibitem{zamora}
\bibinfo{author}{\bibfnamefont{A.}~\bibnamefont{Zamora}},
  \bibinfo{author}{\bibfnamefont{G.}~\bibnamefont{Dagvadorj}},
  \bibinfo{author}{\bibfnamefont{P.}~\bibnamefont{Comaron}},
  \bibinfo{author}{\bibfnamefont{I.}~\bibnamefont{Carusotto}},
  \bibinfo{author}{\bibfnamefont{N.~P.} \bibnamefont{Proukakis}},
  \bibnamefont{and} \bibinfo{author}{\bibfnamefont{M.~H.}
  \bibnamefont{Szyma\ifmmode~\acute{n}\else \'{n}\fi{}ska}},
  \emph{\bibinfo{title}{Kibble-zurek mechanism in driven dissipative systems
  crossing a nonequilibrium phase transition}}, \bibinfo{journal}{Phys. Rev.
  Lett.} \textbf{\bibinfo{volume}{125}}, \bibinfo{pages}{095301}
  (\bibinfo{year}{2020}).

\bibitem{larson}
\bibinfo{author}{\bibfnamefont{P.}~\bibnamefont{Hedvall}} \bibnamefont{and}
  \bibinfo{author}{\bibfnamefont{J.}~\bibnamefont{Larson}},
  \emph{\bibinfo{title}{Dynamics of non-equilibrium steady state quantum phase
  transitions}}, \bibinfo{note}{arXiv:1712.01560}.

\bibitem{patane}
\bibinfo{author}{\bibfnamefont{D.}~\bibnamefont{Patan\`e}},
  \bibinfo{author}{\bibfnamefont{A.}~\bibnamefont{Silva}},
  \bibinfo{author}{\bibfnamefont{L.}~\bibnamefont{Amico}},
  \bibinfo{author}{\bibfnamefont{R.}~\bibnamefont{Fazio}}, \bibnamefont{and}
  \bibinfo{author}{\bibfnamefont{G.~E.} \bibnamefont{Santoro}},
  \emph{\bibinfo{title}{Adiabatic dynamics in open quantum critical many-body
  systems}}, \bibinfo{journal}{Phys. Rev. Lett.}
  \textbf{\bibinfo{volume}{101}}, \bibinfo{pages}{175701}
  (\bibinfo{year}{2008}).

\bibitem{yin2016}
\bibinfo{author}{\bibfnamefont{S.}~\bibnamefont{Yin}},
  \bibinfo{author}{\bibfnamefont{C.-Y.} \bibnamefont{Lo}}, \bibnamefont{and}
  \bibinfo{author}{\bibfnamefont{P.}~\bibnamefont{Chen}},
  \emph{\bibinfo{title}{Scaling in driven dynamics starting in the vicinity of
  a quantum critical point}}, \bibinfo{journal}{Phys. Rev. B}
  \textbf{\bibinfo{volume}{94}}, \bibinfo{pages}{064302}
  (\bibinfo{year}{2016}).

\bibitem{anglin}
\bibinfo{author}{\bibfnamefont{J.~R.} \bibnamefont{Anglin}} \bibnamefont{and}
  \bibinfo{author}{\bibfnamefont{W.~H.} \bibnamefont{Zurek}},
  \emph{\bibinfo{title}{Vortices in the wake of rapid bose-einstein
  condensation}}, \bibinfo{journal}{Phys. Rev. Lett.}
  \textbf{\bibinfo{volume}{83}}, \bibinfo{pages}{1707} (\bibinfo{year}{1999}).

\bibitem{witkowska}
\bibinfo{author}{\bibfnamefont{E.}~\bibnamefont{Witkowska}},
  \bibinfo{author}{\bibfnamefont{P.}~\bibnamefont{Deuar}},
  \bibinfo{author}{\bibfnamefont{M.}~\bibnamefont{Gajda}}, \bibnamefont{and}
  \bibinfo{author}{\bibfnamefont{K.}~\bibnamefont{Rzazewski}},
  \emph{\bibinfo{title}{Solitons as the early stage of quasicondensate
  formation during evaporative cooling}}, \bibinfo{journal}{Phys. Rev. Lett.}
  \textbf{\bibinfo{volume}{106}}, \bibinfo{pages}{135301}
  (\bibinfo{year}{2011}).

\bibitem{Liu2020}
\bibinfo{author}{\bibfnamefont{I.-K.} \bibnamefont{Liu}},
  \bibinfo{author}{\bibfnamefont{J.}~\bibnamefont{Dziarmaga}},
  \bibinfo{author}{\bibfnamefont{S.-C.} \bibnamefont{Gou}},
  \bibinfo{author}{\bibfnamefont{F.}~\bibnamefont{Dalfovo}}, \bibnamefont{and}
  \bibinfo{author}{\bibfnamefont{N.~P.} \bibnamefont{Proukakis}},
  \emph{\bibinfo{title}{Kibble-zurek dynamics in a trapped ultracold bose
  gas}}, \bibinfo{journal}{Phys. Rev. Research} \textbf{\bibinfo{volume}{2}},
  \bibinfo{pages}{033183} (\bibinfo{year}{2020}).

\bibitem{navon}
\bibinfo{author}{\bibfnamefont{N.}~\bibnamefont{Navon}},
  \bibinfo{author}{\bibfnamefont{A.~L.} \bibnamefont{Gaunt}},
  \bibinfo{author}{\bibfnamefont{R.~P.} \bibnamefont{Smith}}, \bibnamefont{and}
  \bibinfo{author}{\bibfnamefont{Z.}~\bibnamefont{Hadzibabic}},
  \emph{\bibinfo{title}{Critical dynamics of spontaneous symmetry breaking in a
  homogeneous bose gas}}, \bibinfo{journal}{Science}
  \textbf{\bibinfo{volume}{347}}(\bibinfo{number}{6218}), \bibinfo{pages}{167}
  (\bibinfo{year}{2015}).

\bibitem{kastner}
\bibinfo{author}{\bibfnamefont{E.~C.} \bibnamefont{King}},
  \bibinfo{author}{\bibfnamefont{J.~N.} \bibnamefont{Kriel}}, \bibnamefont{and}
  \bibinfo{author}{\bibfnamefont{M.}~\bibnamefont{Kastner}},
  \emph{\bibinfo{title}{Universal cooling dynamics toward a quantum critical
  point}}, \bibinfo{journal}{Phys. Rev. Lett.} \textbf{\bibinfo{volume}{130}},
  \bibinfo{pages}{050401} (\bibinfo{year}{2023}).

\bibitem{clerkrmp}
\bibinfo{author}{\bibfnamefont{A.~A.} \bibnamefont{Clerk}},
  \bibinfo{author}{\bibfnamefont{M.~H.} \bibnamefont{Devoret}},
  \bibinfo{author}{\bibfnamefont{S.~M.} \bibnamefont{Girvin}},
  \bibinfo{author}{\bibfnamefont{F.}~\bibnamefont{Marquardt}},
  \bibnamefont{and} \bibinfo{author}{\bibfnamefont{R.~J.}
  \bibnamefont{Schoelkopf}}, \emph{\bibinfo{title}{Introduction to quantum
  noise, measurement, and amplification}}, \bibinfo{journal}{Rev. Mod. Phys.}
  \textbf{\bibinfo{volume}{82}}, \bibinfo{pages}{1155} (\bibinfo{year}{2010}).

\bibitem{sachdev}
\bibinfo{author}{\bibfnamefont{S.}~\bibnamefont{Sachdev}},
  \emph{\bibinfo{title}{Quantum Phase Transitions}}
  (\bibinfo{publisher}{Cambridge Univ. Press}, \bibinfo{address}{Cambridge},
  \bibinfo{year}{1999}).

\bibitem{cardy}
\bibinfo{author}{\bibfnamefont{J.}~\bibnamefont{Cardy}},
  \emph{\bibinfo{title}{Scaling and Renormalization in Statistical Physics}}
  (\bibinfo{publisher}{Cambridge University Press},
  \bibinfo{address}{Cambridge}, \bibinfo{year}{1996}).

\bibitem{continentino}
\bibinfo{author}{\bibfnamefont{M.}~\bibnamefont{Continentino}},
  \emph{\bibinfo{title}{Quantum Scaling in Many-Body Systems: An Approach to
  Quantum Phase Transitions}} (\bibinfo{publisher}{Cambridge University Press},
  \bibinfo{year}{2017}), \bibinfo{edition}{2nd} ed.

\bibitem{brenes}
\bibinfo{author}{\bibfnamefont{M.}~\bibnamefont{Brenes}},
  \bibinfo{author}{\bibfnamefont{J.~J.} \bibnamefont{Mendoza-Arenas}},
  \bibinfo{author}{\bibfnamefont{A.}~\bibnamefont{Purkayastha}},
  \bibinfo{author}{\bibfnamefont{M.~T.} \bibnamefont{Mitchison}},
  \bibinfo{author}{\bibfnamefont{S.~R.} \bibnamefont{Clark}}, \bibnamefont{and}
  \bibinfo{author}{\bibfnamefont{J.}~\bibnamefont{Goold}},
  \emph{\bibinfo{title}{Tensor-network method to simulate strongly interacting
  quantum thermal machines}}, \bibinfo{journal}{Phys. Rev. X}
  \textbf{\bibinfo{volume}{10}}, \bibinfo{pages}{031040}
  (\bibinfo{year}{2020}).

\bibitem{weiss}
\bibinfo{author}{\bibfnamefont{U.}~\bibnamefont{Weiss}},
  \emph{\bibinfo{title}{Quantum Dissipative Systems}}
  (\bibinfo{publisher}{World Scientific}, \bibinfo{address}{Singapore},
  \bibinfo{year}{2000}).

\bibitem{herbutbook}
\bibinfo{author}{\bibfnamefont{I.}~\bibnamefont{Herbut}},
  \emph{\bibinfo{title}{A Modern Approach to Critical Phenomena}}
  (\bibinfo{publisher}{Cambridge University Press}, \bibinfo{year}{2007}).

\bibitem{polkovnikov2008}
\bibinfo{author}{\bibfnamefont{R.}~\bibnamefont{Barankov}} \bibnamefont{and}
  \bibinfo{author}{\bibfnamefont{A.}~\bibnamefont{Polkovnikov}},
  \emph{\bibinfo{title}{Optimal nonlinear passage through a quantum critical
  point}}, \bibinfo{journal}{Phys. Rev. Lett.} \textbf{\bibinfo{volume}{101}},
  \bibinfo{pages}{076801} (\bibinfo{year}{2008}).

\bibitem{sen2008}
\bibinfo{author}{\bibfnamefont{D.}~\bibnamefont{Sen}},
  \bibinfo{author}{\bibfnamefont{K.}~\bibnamefont{Sengupta}}, \bibnamefont{and}
  \bibinfo{author}{\bibfnamefont{S.}~\bibnamefont{Mondal}},
  \emph{\bibinfo{title}{Defect production in nonlinear quench across a quantum
  critical point}}, \bibinfo{journal}{Phys. Rev. Lett.}
  \textbf{\bibinfo{volume}{101}}, \bibinfo{pages}{016806}
  (\bibinfo{year}{2008}).

\bibitem{ROSSINI2021}
\bibinfo{author}{\bibfnamefont{D.}~\bibnamefont{Rossini}} \bibnamefont{and}
  \bibinfo{author}{\bibfnamefont{E.}~\bibnamefont{Vicari}},
  \emph{\bibinfo{title}{Coherent and dissipative dynamics at quantum phase
  transitions}}, \bibinfo{journal}{Physics Reports}
  \textbf{\bibinfo{volume}{936}}, \bibinfo{pages}{1} (\bibinfo{year}{2021}),
  \bibinfo{note}{coherent and dissipative dynamics at quantum phase
  transitions}.

\bibitem{dutta2015}
\bibinfo{author}{\bibfnamefont{A.}~\bibnamefont{Dutta}},
  \bibinfo{author}{\bibfnamefont{G.}~\bibnamefont{Aeppli}},
  \bibinfo{author}{\bibfnamefont{B.~K.} \bibnamefont{Chakrabarti}},
  \bibinfo{author}{\bibfnamefont{U.}~\bibnamefont{Divakaran}},
  \bibinfo{author}{\bibfnamefont{T.~F.} \bibnamefont{Rosenbaum}},
  \bibnamefont{and} \bibinfo{author}{\bibfnamefont{D.}~\bibnamefont{Sen}},
  \emph{\bibinfo{title}{Quantum Phase Transitions in Transverse Field Spin
  Models: From Statistical Physics to Quantum Information}}
  (\bibinfo{publisher}{Cambridge University Press}, \bibinfo{year}{2015}).

\bibitem{coldea}
\bibinfo{author}{\bibfnamefont{R.}~\bibnamefont{Coldea}},
  \bibinfo{author}{\bibfnamefont{D.~A.} \bibnamefont{Tennant}},
  \bibinfo{author}{\bibfnamefont{E.~M.} \bibnamefont{Wheeler}},
  \bibinfo{author}{\bibfnamefont{E.}~\bibnamefont{Wawrzynska}},
  \bibinfo{author}{\bibfnamefont{D.}~\bibnamefont{Prabhakaran}},
  \bibinfo{author}{\bibfnamefont{M.}~\bibnamefont{Telling}},
  \bibinfo{author}{\bibfnamefont{K.}~\bibnamefont{Habicht}},
  \bibinfo{author}{\bibfnamefont{P.}~\bibnamefont{Smeibidl}}, \bibnamefont{and}
  \bibinfo{author}{\bibfnamefont{K.}~\bibnamefont{Kiefer}},
  \emph{\bibinfo{title}{Quantum criticality in an ising chain: Experimental
  evidence for emergent $e_8$ symmetry}}, \bibinfo{journal}{Science}
  \textbf{\bibinfo{volume}{327}}, \bibinfo{pages}{177} (\bibinfo{year}{2010}).

\bibitem{kinross}
\bibinfo{author}{\bibfnamefont{A.~W.} \bibnamefont{Kinross}},
  \bibinfo{author}{\bibfnamefont{M.}~\bibnamefont{Fu}},
  \bibinfo{author}{\bibfnamefont{T.~J.} \bibnamefont{Munsie}},
  \bibinfo{author}{\bibfnamefont{H.~A.} \bibnamefont{Dabkowska}},
  \bibinfo{author}{\bibfnamefont{G.~M.} \bibnamefont{Luke}},
  \bibinfo{author}{\bibfnamefont{S.}~\bibnamefont{Sachdev}}, \bibnamefont{and}
  \bibinfo{author}{\bibfnamefont{T.}~\bibnamefont{Imai}},
  \emph{\bibinfo{title}{Evolution of quantum fluctuations near the quantum
  critical point of the transverse field ising chain system
  ${\mathrm{conb}}_{2}{\mathrm{o}}_{6}$}}, \bibinfo{journal}{Phys. Rev. X}
  \textbf{\bibinfo{volume}{4}}, \bibinfo{pages}{031008} (\bibinfo{year}{2014}).

\bibitem{King2022}
\bibinfo{author}{\bibfnamefont{A.~D.} \bibnamefont{King}},
  \bibinfo{author}{\bibfnamefont{S.}~\bibnamefont{Suzuki}},
  \bibinfo{author}{\bibfnamefont{J.}~\bibnamefont{Raymond}},
  \bibinfo{author}{\bibfnamefont{A.}~\bibnamefont{Zucca}},
  \bibinfo{author}{\bibfnamefont{T.}~\bibnamefont{Lanting}},
  \bibinfo{author}{\bibfnamefont{F.}~\bibnamefont{Altomare}},
  \bibinfo{author}{\bibfnamefont{A.~J.} \bibnamefont{Berkley}},
  \bibinfo{author}{\bibfnamefont{S.}~\bibnamefont{Ejtemaee}},
  \bibinfo{author}{\bibfnamefont{E.}~\bibnamefont{Hoskinson}},
  \bibinfo{author}{\bibfnamefont{S.}~\bibnamefont{Huang}},
  \bibinfo{author}{\bibfnamefont{E.}~\bibnamefont{Ladizinsky}},
  \bibinfo{author}{\bibfnamefont{A.~J.~R.} \bibnamefont{MacDonald}},
  \emph{et~al.}, \emph{\bibinfo{title}{Coherent quantum annealing in a
  programmable 2,000{\thinspace}qubit ising chain}}, \bibinfo{journal}{Nature
  Physics}  (\bibinfo{year}{2022}).

\bibitem{Bando2020}
\bibinfo{author}{\bibfnamefont{Y.}~\bibnamefont{Bando}},
  \bibinfo{author}{\bibfnamefont{Y.}~\bibnamefont{Susa}},
  \bibinfo{author}{\bibfnamefont{H.}~\bibnamefont{Oshiyama}},
  \bibinfo{author}{\bibfnamefont{N.}~\bibnamefont{Shibata}},
  \bibinfo{author}{\bibfnamefont{M.}~\bibnamefont{Ohzeki}},
  \bibinfo{author}{\bibfnamefont{F.~J.} \bibnamefont{G\'omez-Ruiz}},
  \bibinfo{author}{\bibfnamefont{D.~A.} \bibnamefont{Lidar}},
  \bibinfo{author}{\bibfnamefont{S.}~\bibnamefont{Suzuki}},
  \bibinfo{author}{\bibfnamefont{A.}~\bibnamefont{del Campo}},
  \bibnamefont{and}
  \bibinfo{author}{\bibfnamefont{H.}~\bibnamefont{Nishimori}},
  \emph{\bibinfo{title}{Probing the universality of topological defect
  formation in a quantum annealer: Kibble-zurek mechanism and beyond}},
  \bibinfo{journal}{Phys. Rev. Research} \textbf{\bibinfo{volume}{2}},
  \bibinfo{pages}{033369} (\bibinfo{year}{2020}).

\bibitem{daley}
\bibinfo{author}{\bibfnamefont{A.~J.} \bibnamefont{Daley}},
  \emph{\bibinfo{title}{Quantum trajectories and open many-body quantum
  systems}}, \bibinfo{journal}{Advances in Physics}
  \textbf{\bibinfo{volume}{63}}, \bibinfo{pages}{77} (\bibinfo{year}{2014}).

\bibitem{abbruzzo2021}
\bibinfo{author}{\bibfnamefont{A.}~\bibnamefont{D'Abbruzzo}} \bibnamefont{and}
  \bibinfo{author}{\bibfnamefont{D.}~\bibnamefont{Rossini}},
  \emph{\bibinfo{title}{Self-consistent microscopic derivation of markovian
  master equations for open quadratic quantum systems}},
  \bibinfo{journal}{Phys. Rev. A} \textbf{\bibinfo{volume}{103}},
  \bibinfo{pages}{052209} (\bibinfo{year}{2021}).

\bibitem{Laine2012}
\bibinfo{author}{\bibfnamefont{E.-M.} \bibnamefont{Laine}},
  \bibinfo{author}{\bibfnamefont{K.}~\bibnamefont{Luoma}}, \bibnamefont{and}
  \bibinfo{author}{\bibfnamefont{J.}~\bibnamefont{Piilo}},
  \emph{\bibinfo{title}{Local-in-time master equations with memory effects:
  applicability and interpretation}}, \bibinfo{journal}{Journal of Physics B:
  Atomic, Molecular and Optical Physics}
  \textbf{\bibinfo{volume}{45}}(\bibinfo{number}{15}), \bibinfo{pages}{154004}
  (\bibinfo{year}{2012}).

\bibitem{breuertimelocal2019}
\bibinfo{author}{\bibfnamefont{G.}~\bibnamefont{Amato}},
  \bibinfo{author}{\bibfnamefont{H.-P.} \bibnamefont{Breuer}},
  \bibnamefont{and} \bibinfo{author}{\bibfnamefont{B.}~\bibnamefont{Vacchini}},
  \emph{\bibinfo{title}{Microscopic modeling of general time-dependent quantum
  markov processes}}, \bibinfo{journal}{Phys. Rev. A}
  \textbf{\bibinfo{volume}{99}}, \bibinfo{pages}{030102}
  (\bibinfo{year}{2019}).

\bibitem{Donvil2022}
\bibinfo{author}{\bibfnamefont{B.}~\bibnamefont{Donvil}} \bibnamefont{and}
  \bibinfo{author}{\bibfnamefont{P.}~\bibnamefont{Muratore-Ginanneschi}},
  \emph{\bibinfo{title}{Quantum trajectory framework for general time-local
  master equations}}, \bibinfo{journal}{Nature Communications}
  \textbf{\bibinfo{volume}{13}}(\bibinfo{number}{1}), \bibinfo{pages}{4140}
  (\bibinfo{year}{2022}).

\bibitem{deffner17}
\bibinfo{author}{\bibfnamefont{S.}~\bibnamefont{Deffner}},
  \emph{\bibinfo{title}{Kibble-zurek scaling of the irreversible entropy
  production}}, \bibinfo{journal}{Phys. Rev. E} \textbf{\bibinfo{volume}{96}},
  \bibinfo{pages}{052125} (\bibinfo{year}{2017}).

\bibitem{liang}
\bibinfo{author}{\bibfnamefont{T.}~\bibnamefont{Liang}},
  \bibinfo{author}{\bibfnamefont{S.~M.} \bibnamefont{Koohpayeh}},
  \bibinfo{author}{\bibfnamefont{J.~W.} \bibnamefont{Krizan}},
  \bibinfo{author}{\bibfnamefont{T.~M.} \bibnamefont{McQueen}},
  \bibinfo{author}{\bibfnamefont{R.~J.} \bibnamefont{Cava}}, \bibnamefont{and}
  \bibinfo{author}{\bibfnamefont{N.~P.} \bibnamefont{Ong}},
  \emph{\bibinfo{title}{Heat capacity peak at the quantum critical point of the
  transverse ising magnet conb2o6}}, \bibinfo{journal}{Nature Communications}
  \textbf{\bibinfo{volume}{6}}, \bibinfo{pages}{7611} (\bibinfo{year}{2015}).

\bibitem{Alicki2018}
\bibinfo{author}{\bibfnamefont{R.}~\bibnamefont{Alicki}} \bibnamefont{and}
  \bibinfo{author}{\bibfnamefont{R.}~\bibnamefont{Kosloff}},
  \emph{\bibinfo{title}{Introduction to Quantum Thermodynamics: History and
  Prospects}} (\bibinfo{publisher}{Springer International Publishing},
  \bibinfo{address}{Cham}, \bibinfo{year}{2018}), pp. \bibinfo{pages}{1--33},
  ISBN \bibinfo{isbn}{978-3-319-99046-0}.

\bibitem{gluza}
\bibinfo{author}{\bibfnamefont{M.}~\bibnamefont{Gluza}},
  \bibinfo{author}{\bibfnamefont{J.}~\bibnamefont{Sabino}},
  \bibinfo{author}{\bibfnamefont{N.~H.} \bibnamefont{Ng}},
  \bibinfo{author}{\bibfnamefont{G.}~\bibnamefont{Vitagliano}},
  \bibinfo{author}{\bibfnamefont{M.}~\bibnamefont{Pezzutto}},
  \bibinfo{author}{\bibfnamefont{Y.}~\bibnamefont{Omar}},
  \bibinfo{author}{\bibfnamefont{I.}~\bibnamefont{Mazets}},
  \bibinfo{author}{\bibfnamefont{M.}~\bibnamefont{Huber}},
  \bibinfo{author}{\bibfnamefont{J.}~\bibnamefont{Schmiedmayer}},
  \bibnamefont{and} \bibinfo{author}{\bibfnamefont{J.}~\bibnamefont{Eisert}},
  \emph{\bibinfo{title}{Quantum field thermal machines}}, \bibinfo{journal}{PRX
  Quantum} \textbf{\bibinfo{volume}{2}}, \bibinfo{pages}{030310}
  (\bibinfo{year}{2021}).

\bibitem{nielsen}
\bibinfo{author}{\bibfnamefont{M.}~\bibnamefont{Nielsen}} \bibnamefont{and}
  \bibinfo{author}{\bibfnamefont{I.}~\bibnamefont{Chuang}},
  \emph{\bibinfo{title}{Quantum Computation and Quantum Information}}
  (\bibinfo{publisher}{Cambridge University Press},
  \bibinfo{address}{Cambridge}, \bibinfo{year}{2000}).

\end{thebibliography}

\newpage
\appendix



\section{Methods}
\subsection{Diagonalization of the Hamiltonian in the transverse field Ising model}

With the Jordan-Wigner transformation $\sigma_j^z = \left(c_j + c_j^+ \right)\prod_{m<j}e^{i\pi \hat{n}_m}$ and $\sigma_j^x = 1 - 2\hat{n}_j$
with $\hat{n}_j= c_j^+c_j$ as introduced in Ref. \cite{dziarmaga}, the Hamiltonian reads

\begin{gather}
H= - J\sum_{j}\left[ g(1-2c_j^+c_j) + \left(c_j^+ c_{j+1} + c_j^+ c_{j+1}^+ + h.c.\right)\right]
\end{gather}

By applying Fourier transform to momentum space, we obtain

\begin{gather}
H = 
 -NJg + 2J\sum_{k} \left[ A_k c_k^+ c_k +  \left( \frac{iB_k}{2} c_{-k}c_{k} + h.c.\right)\right]
\end{gather}
where $A_k = g-\cos(ka)$ and $B_k = \sin(ka)$. 

The Hamiltonian is diagonalized by the Bogoliubov transformation 

\begin{gather}
c_k = u_k d_{k+} + v_k^* d_{k-}^+ \nonumber \\
c_{-k} = u_k d_{k-} - v_k^* d_{k+}^+
\end{gather}
where 
$u_k = \sqrt{1 + (1+B_k/A_k)^{-1/2}}/\sqrt{2}$ and $v_k = -i\sqrt{1 - (1+B_k/A_k)^{-1/2}}/\sqrt{2}$.
After Bogoliubov transformation the Hamiltonian reads

\begin{gather}
H = \sum_{k>0,m=\pm}E_k \left(d_{km}^+d_{km}-\frac 12\right)
\end{gather}
where 
$E_k = 2J\sqrt{(g-\cos(ka))^2 + \sin^2(ka)}$
is the energy spectrum of the fermionic excitations. The energy dependent density of states $G(E)$ is calculated based on the fact that the density of states should be preserves both in momentum and energy space, expressed as  $G(E)\mathrm{d}E = 2\frac{L}{2\pi}dk$ where the factor $2$ stems from the $m$-degeneracy. Substituting the spectrum and expressing the wavenumber with the energy leads to
\begin{gather}
G(E)=\frac{L}{\pi}\left|\frac{\mathrm{d}E}{\mathrm{d}k}\right|^{-1}=\frac{N}{2J}\frac{1}{\sqrt{2\left(g^2+1\right) - \left(g^2-1\right)^2\left(\frac{2J}{E}\right)^2 - \left(\frac{E}{2J}\right)^2 }}\,.
\end{gather}

\subsection{Derivation of the number of defects after temperature quench in the transverse field Ising model}
In this section, the derivation of the density of defects is presented.
The number of defects is defined as $\mathcal{N}_S(t) = \sum_{km} p_k(t)$ where $p_k(t) = \langle d_{km}^+d_{km} \rangle(t)$ is the occupation probability of the fermionic state corresponding to the quantum numbers $k$ and $m$.

In order to determine the dynamics of $p_k(t)$, let us recall the Lindblad equation

\begin{gather}
\partial_t\rho = -i\left[ H,\rho\right] 
+ \sum_{km}\gamma_{k,\uparrow}(t) \mathcal{D}\left(L_{km,\uparrow};\rho; L_{km,\uparrow}^+\right) +
\gamma_{k,\downarrow}(t) \mathcal{D}\left(L_{km,\downarrow};\rho; L_{km,\downarrow}^+\right)
\end{gather}
where $L_{km,\uparrow} = d_{km}^+$, $L_{km,\downarrow} = d_{km}$, $H=\sum_{k>0,m}E_k (d_{km}^+d_{km}-\frac 12)$ and the coupling constants are given by

\begin{gather}
\gamma_{k,\uparrow}(t) = \gamma \frac{1}{1 + e^{\beta(t) E_k}}
\end{gather}
and $\gamma_{k,\downarrow}(t) = \gamma - \gamma_{k,\uparrow}$
with the time-dependent temperature $T(t) = T_0(1-t/\tau)$. Note that both the unitary and the dissipative terms are diagonal in $k$ and $m$. Hence, for each $km$, the dynamics is restricted to a two-dimensional Hilbert space spanned by the states that $km$ is empty or occupied. Using the empty and occupied states as basis, the density matrix can be represented by
\begin{gather}
\rho(t) =\prod_{k>0,m} \left[\begin{array}{cc} 1-p_k(t) & q_k(t) \\ q_k(t)^* & p_k(t) \end{array}\right]
\end{gather}
with the real-valued functions $p_k(t) = \langle d_{km}^+ d_{km}\rangle $ and complex-valued functions $q_k(t)$.
Based on the Lindblad equation, the following equations are derived.

\begin{subequations}
\begin{gather}
\dot{p}_k = \gamma_{k,\uparrow}(t)(1-p_k(t)) - \gamma_{k,\downarrow}(t) p_k(t) \\
\dot{q}_k = \left( i\Delta - \frac{\gamma_{k,\uparrow}(t) + \gamma_{k,\downarrow}(t)}{2}\right) q_k(t)
\end{gather}
\label{eq:pqt}
\end{subequations}
If $\gamma_{k,\uparrow}$ and $\gamma_{k,\downarrow}$ did not change with time, the steady state of Eq. \eqref{eq:pqt} would be $q_{k,\infty}=0$ and $p_{k,\infty} = (1+\gamma_{k,\downarrow}/\gamma_{k,\uparrow})^{-1} = (1 + e^{\beta E_k})^{-1}$ describing thermal equilibrium.

\begin{figure*}[t]
\centering
\includegraphics[width=18cm]{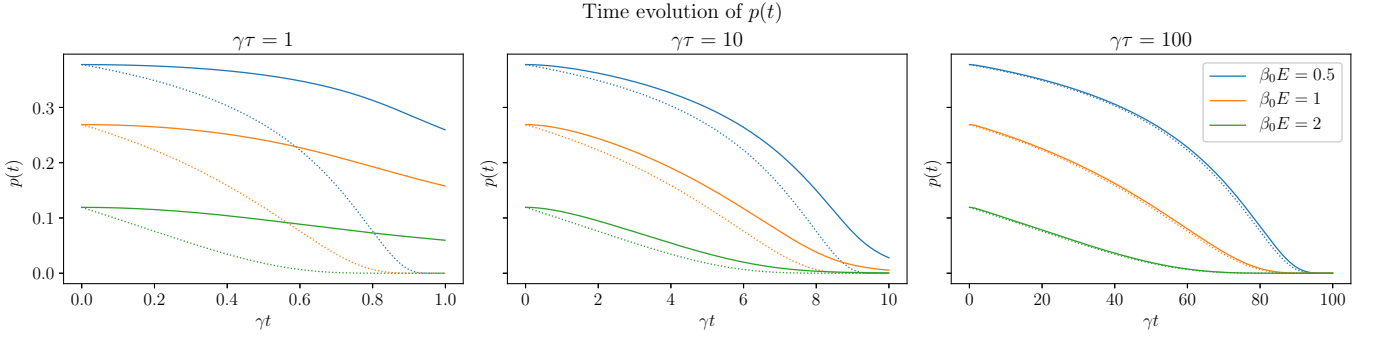}
\caption{The time evolution of occupation probability for different quench duration. The solid line is the numerical solution of Eq. \eqref{eq:tdep} while the dashed line corresponds to the system in thermal equilibrium at the instantaneous temperature. }
\label{fig:poft}
\end{figure*}

In our model, we assume that the initial condition is the thermal equilibrium state corresponding to the initial temperature $T_0$, i.e., $p(0) = (1+e^{\beta_0 E})^{-1}$ with $\beta_0 = T_0^{-1}$ and $q(0)=0$. The inhomogeneous differential equations in Eqs. \eqref{eq:pqt} are solved by

\begin{gather}
p_k(t) = \frac{e^{-\gamma t}}{1+e^{\beta_0 E_k}} + \gamma\int_0^{t} \frac{e^{-\gamma(t-t')}}{1+e^{\beta(t')E_k}} \mathrm{d}t' 
\label{eq:tdep}
\end{gather}
and $q_k(t)=0$ for the time interval $0<t<\tau$. After the temperature quench, i.e., when the cooling has already ended and the temperature is constant zero, we obtain $p(t>\tau) = p(\tau) e^{-\gamma (t-\tau)}$.

The time dependence of $p(t)$ is evaluated numerically and is shown in Fig. \ref{fig:poft} for several quench durations. In the figure, dashed lines show the probability for the system in thermal equilibrium at the instantaneous temperature. 
It can be observed that for long quenches, the time evolution follows closely the equilibrium values but for short quenches, they differ significantly.

The final number of defects is obtained by summing up Eq. \eqref{eq:tdep} leading to

\begin{gather}
\mathcal{N}_S(t) = 
F(\beta_0) e^{-\gamma t} + \gamma\int_0^t e^{-\gamma (t-t')}F(\beta(t')) \mathrm{d}t' 
\label{eq:Nst}
\end{gather}
where

\begin{gather}
F(\beta)= \int_{\Delta}^{2J|g+1|} \mathrm{d}E\frac{G(E)}{1 + e^{\beta E}}
\label{eq:Fb}
\end{gather}
is the expectation value of the total number of fermionic excitations in the system where $\Delta = 2J|g-1|$ is the gap.

We are mostly interested in the low-temperature behavior, i.e., when the temperature is much lower than the bandwidth during the whole quench, $T_0\ll 2J$. In this situation, only low-energy states are occupied for which the density of states is approximated as

\begin{gather}
G(E)\approx \frac{N}{2\pi J} \left\{\begin{array}{cc} \sqrt{\frac{E}{2g(E-\Delta)}} & \mbox{if $g$ is far from 1} \\
1 & \mbox{if $g=1$} \end{array}\right.
\label{eq:gapge}
\end{gather}
and the function $F(\beta)$ is computed as

\begin{gather}
\frac{F(\beta)}{N} \approx\left\{\begin{array}{cc} \sqrt{\frac{|g-1|}{4\pi g  \beta J}}e^{-\beta \Delta} & \mbox{if $g$ is far from 1} \\
\frac{\ln 2}{2\pi\beta J} & \mbox{if $g=1$} \end{array}\right.
\label{eq:Fg}
\end{gather}
where the upper limit of the integral in Eq. \eqref{eq:Fb} has been set to infinity.

Numerical investigations show that the approximate functions in Eq. \eqref{eq:Fg} are in good agreement with the numerically evaluated Eq. \eqref{eq:Fb} at low temperatures.
Substituting Eqs. \eqref{eq:Fg} into Eq. \eqref{eq:Nst}, we obtain

\begin{gather}
n(t) = \frac{\mathcal{N}_S(t)}{N} = \sqrt{\frac{|1-g|}{4\pi J\beta_0 g}\left(1-\frac{t}{\tau}\right)} e^{-\frac{\beta_0\Delta}{1-\frac{t}{\tau}}} + e^{-\gamma t}\sqrt{\frac{|1-g|}{ 64J\beta_0 g}}e^{-\beta_0\Delta}  \sum_{j=\pm 1} e^{\left(\sqrt{\beta_0\Delta} + j\sqrt{\gamma\tau}\right)^2}\times\nonumber \\
\times\left(2\sqrt{\beta_0\Delta} - \frac{j}{\sqrt{\gamma\tau}}\right) \left[\Phi\left(\frac{\sqrt{\beta_0\Delta}}{\sqrt{1-\frac{t}{\tau}}}+ j \sqrt{\gamma\tau}\sqrt{1-\frac{t}{\tau}}\right) - \Phi\left(\sqrt{\beta_0\Delta} + j\sqrt{\gamma\tau}\right)\right]
\end{gather}
if $g$ is far from 1. \jav{In the formula,
$\Phi(x)$ is the error function defined as $\Phi(x) = \frac{2}{\sqrt{\pi}}\int_{0}^x e^{-y^2}\mathrm{d}y$. If $g=1$,}
\begin{gather}
n(t) = \frac{\ln 2}{2\pi J\beta_0} 
\left(1-\frac{t}{\tau}   + \frac{1 -e^{-\gamma t}}{\gamma\tau}\right)\,.
\end{gather}
At the end of the quench, $t=\tau$, the density of defects is calculated as
\begin{gather}
n(\tau) = \sqrt{\frac{|1-g|}{64J\beta_0 g}} \sum_{j=\pm 1} e^{2j\sqrt{\beta_0\Delta\gamma\tau}} \left(2\sqrt{\beta_0\Delta} - \frac{j}{\sqrt{\gamma\tau}}\right) \left[1 - \Phi\left(\sqrt{\beta_0\Delta} + j\sqrt{\gamma\tau}\right)\right]
\label{eq:nonunit}
\end{gather}
for $g$ being far from 1 and
\begin{gather}
n(\tau) = \frac{\ln 2}{2\pi J\beta_0} \frac{1 - e^{-\gamma \tau}}{\gamma\tau}
\label{eq:unit}
\end{gather}
for $g=1$.

\subsection{Defect density in the transverse field Ising model coupled to an ohmic thermal bath}
In this section, we assume that the transverse field Ising chain is coupled to an environment with the coupling constants

\begin{gather}
\gamma_{k,\uparrow}(t) =  \gamma(E_k) \frac{1}{1 + e^{\beta(t) E_k}}
\\
\gamma_{k,\downarrow}(t) = \gamma(E_k) \frac{e^{\beta(t)E_k}}{1 + e^{\beta(t) E_k}}
\end{gather}
with $\gamma(E)=\gamma_0 \frac{E}{2J}$ and $E_k$ is the positive-valued spectrum of fermionic excitations.
In contrast to the previous model, the coupling constants are characterized by an effective spectral density proportional to the energy as typical for Ohmic environment \cite{weiss}. The normalization with $2J$ has been introduced to preserve the dimension of $\gamma_0$.

\begin{figure}[h]
\centering
\includegraphics[width=8cm]{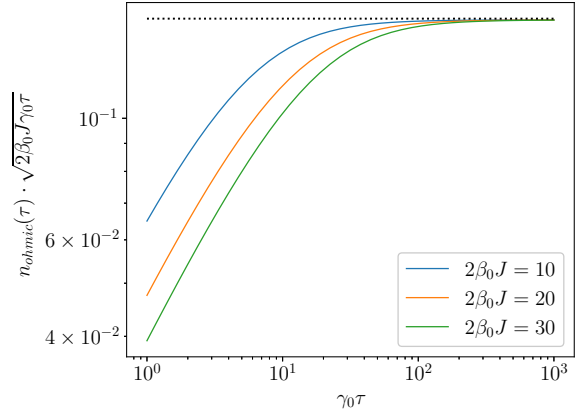}
\caption{Final defect density at $g=1$ in the case of Ohmic environment obtained by numerically integrating Eq. \eqref{eq:ohmicn}. For large $\tau$, the combination $n(\tau)\sqrt{2\beta_0 J\gamma_0\tau}$ tends to a constant value which equals approximately $0.152$ implying that $n(\tau)\sim 1/\sqrt{\tau}$.}
\label{fig:ohmicn}
\end{figure}

The time evolution is formally the same in each wavenumber sector as before in Eq. \eqref{eq:tdep} but the relaxation rate $\gamma$ should be replaced by $\gamma_0 E_k$. The final density of defects is determined by the integral

\begin{gather}
n(\tau)= 
\int_0^\infty \frac{G(E)}{N}\left(\frac{e^{-\frac{\gamma_0 E}{2J} \tau}}{1+e^{\beta_0 E}} + \frac{\gamma_0 E}{2J}\int_0^{\tau} \frac{e^{-\frac{\gamma_0 E}{2J}(\tau-t')}}{1+e^{\beta(t')E}} \mathrm{d}t'  \right)\mathrm{d}E
\label{eq:ohmicn}
\end{gather}
In the $g=1$ case, the defect density is expected to scale as $n(\tau)\sim \tau^{-\frac{d}{z(1+s)}}$ where $d=1$ is the dimension of the transverse field Ising model, $z=1$ is the dynamical critical exponent and $s=1$ for Ohmic environment. Hence, $n(\tau)\sim 1/\sqrt{\tau}$.
The scaling law is confirmed by numerically evaluating the integrals in Eq. \eqref{eq:ohmicn}. As shown in Fig. \ref{fig:ohmicn}, the combination of $n(\tau)\sqrt{\tau}$ converges to a constant value indicating that the defect density obeys $n(\tau)\sim 1/\sqrt{\tau}$ indeed.

\begin{figure}[h]
\centering
\includegraphics[width=8cm]{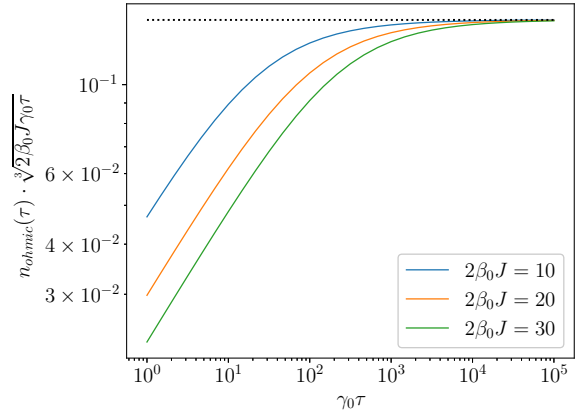}
\caption{Final defect density at $g=1$ in the case of non-Ohmic ($s=2$) environment obtained numerically. For large $\tau$, the combination $n(\tau)\cdot\left(2\beta_0 J\gamma_0\tau\right)^{1/3}$ tends to a constant value which equals approximately $0.145$ implying that $n(\tau)\sim \tau^{-1/3}$.}
\label{fig:nonohmicn}
\end{figure}

We note that the scaling law of $n(\tau)\sim \tau^{-\frac{d}{z(1+s)}}$ can also be verified numerically in the case of a non-Ohmic effective spectral density, $\gamma(E) = \gamma_0 (E/2J)^s$. For instance, for $s=2$, $d=z=1$, the scaling law predicts $n(\tau)\sim\tau^{-1/3}$ which is numerically confirmed as shown in Fig. \ref{fig:nonohmicn}.

\end{document}